\documentstyle[epsfig,aps,preprint]{revtex}

\newlength{\dinwidth}
\newlength{\dinmargin}
\def\bea{\begin{eqnarray}}
\def\eea{\end{eqnarray}}
\def\nn{\nonumber}

\begin{document}

\preprint{\bf   hep-ph/0102168}

\title{Possible Supersymmetric Effects on Angular Distributions   \\
in $B \to K^* (\to K \pi) \ell^+ \ell^- $ Decays }

\author{
C.S. Kim${~^a}$\footnote{e-mail: kim@kimcs.yonsei.ac.kr,
~~ http://phya.yonsei.ac.kr/\~{}cskim/}, ~~
Yeong Gyun Kim${~^b}$\footnote{e-mail: ygkim@yukawa.kyoto-u.ac.jp} ~~
and ~~ Cai-Dian L\"u${~^{c,d}}$\footnote{e-mail: 
lucd@hptc5.ihep.ac.cn} }

\address{
$a:$ Department of Physics and IPAP, Yonsei University, Seoul 120-749, Korea \\
$b:$ YITP, Kyoto University, Kyoto 606-8502, Japan \\
$c:$ CCAST (World Lab.), P.O. Box 8730, Beijing 100080,
China \\
$d:$ Theory divison, Institute of High Energy Physics, P.O.Box 918(4),
Beijing 100039,China\footnote{Mailing address} \\
}


\maketitle

\begin{abstract}

\noindent 
We investigate the angular distributions of the rare $B$ decay,
$B \to K^* (\to K \pi) \ell^+ \ell^-$,  
in general supersymmetric extensions of the standard model.
We consider the new physics contributions from the operators $O_{7,8,9,10}$
in small invariant mass region of lepton pair.
We show that the azimuthal angle distribution of the
decay can tell us the new physics effects clearly from the behavior of the
distribution, even if new physics does not change the decay rate
substantially from the standard model prediction.

\end{abstract}



\section{Introduction}

Rare decays of $B$ meson, $e.g.$ $b\to s \gamma$ and
$b\to s \ell^+ \ell^-$,
are the most suitable candidates for study of new
physics beyond the standard model (SM).
Since their branching fractions are usually very small within the SM
predictions,  they can sometimes show up new physics with unexpectedly 
large values of decay rates. 
The decay $b\to s \gamma$, which has been already measured by the CLEO
Collaboration \cite{cleo}, has shown that there is not much extra
parameter space  for its branching fraction from new physics \cite{bur}.
The decay $b\to s \ell^+\ell^-$ suffers from severe backgrounds
of $J/\psi$ and $\psi '$  resonance contributions in the measurement of
branching fraction, 
and so it may not be easy to uncover new physics cleanly \cite{res}.  
However, a number of methods have been discussed to detect new physics
through the details of the decay process, such as  differential distributions
or polarization effects \cite{pre,cskim,dan}.

Recently a new method utilizing the angular distribution of 
$B\to K^* (\to K \pi) \ell^+\ell^-$ 
has been proposed \cite{kklm}. 
Imagine the decay configuration when 
$K^*$  is emitted to the direction of 
$+z$ and $\gamma^*$ is emitted to the opposite direction in the
rest frame of $B$ meson. Here $\gamma^*$
is  off-shell photon and it further decays into $\ell^+  \ell^-$, 
and $K^*$  subsequently decays into $K  \pi$. If we ignore small
mixture of the longitudinal component, 
the angular momentum of $K^*$ is either $J_z=+1$ or $J_z=-1$, 
and the corresponding production amplitude 
is proportional to $C_{7R}$ or $C_{7L}$, respectively.
Suppose the final $K$ meson is emitted to
the direction of $(\theta_K,~ \phi)$ in the rest frame of $K^*$, where
$\theta_K$ is a polar angle and $\phi$ is an azimuthal  angle
between the decay plane of ($K  \pi$) and the decay plane of ($\ell^+  \ell^-$).
In the low invariant $m_{\ell^+ \ell^-}$
region, electromagnetic operator terms are dominated and the decay amplitude
for the whole process is proportional to
$$ 
A~ C_{7L} \exp( -i  \phi) + B~ C_{7R} \exp(+i \phi)  + C~,
$$ 
where $A$, $B$ and $C$ are the real functions of the other angles. 

In this new method, we can distinguish the new physics contribution 
from that of the SM even if the branching fraction of the decay is
similar to the prediction of the SM:
In the $B \to K^* \gamma$ decays 
the probability of $B$ meson decaying to left-handed (or right-handed) 
circular polarized $K^*$ is proportional to $|C_{7L}|^2$ (or $|C_{7R}|^2$),
and therefore 
the polarization measurement of $K^*$ and $\gamma$ is useful for
extracting the ratio of ${|C_{7L}|}/{|C_{7R}|}$.
Even though the polarization of high energy real photon cannot be measured
easily, we can still get some useful information through the azimuthal 
angle distribution in the low invariant mass region of dileptons for 
$B\to K^* (\to K \pi) \ell^+\ell^-$ decay. This is because 
the decay products of $K^*$ and the virtual photon $\gamma^*$ are responsible
for this polarization measurements. 
In the SM, the operator $O_{7L}$ is dominant and the operator $O_{7R}$
is suppressed by ${\cal O} (m_s/m_b)$. In this case, the angular
distribution of the decay products is a flat function of the angle $\phi$
in the small lepton invariant mass region. 
If there is new physics contribution, the 
contribution of both operators can be equally
important \cite{kklm}. 
We can distinguish the new physics signal easily
from the angular distribution of 
the decay $B\to K^* (\to K\pi)\ell^+\ell^-$, 
while the measured branching fraction for $B \to X_s \gamma$ can still be
accommodated.

In this paper, we extend this method to calculate the angular distribution
of $B\to K^* (K\pi) \ell^+\ell^-$ in generalized supersymmetry models
(gSUSYs).  In addition to the operator $O_{7R}$, 
we also consider the new operators $O_{9R}$ and $O_{10R}$. 
In the following section, we derive the general formula for the various
angular distributions including the new operators. 
Section III is devoted to the numerical analysis
and discussions in gSUSYs. 

\section{Angular Distributions of the decay
$B \to K^*(\to K + \pi)+ \ell^+ +\ell^-$}

We start with 
the general effective Hamiltonian for the corresponding 
$b\to s \ell^+\ell^-$ decay \cite{ab},
\bea
{\cal H}_{eff}(b\to s \ell^+\ell^-)&=&
-\frac{G_F}{\sqrt{2}} V_{tb}V_{ts}^* {\alpha \over \pi}
\sum _{i=7}^{10} (C_{iL} O_{iL} + C_{iR} O_{iR})~. 
\eea
The operators $O_i$ relevant for us are
\bea
O_{7L} &=& 
\frac{em_b}{4 \pi^2} (\bar s_L \sigma_{\mu \nu} b_R) F^{\mu\nu},\\
O_{7R} &=& 
\frac{em_b}{4 \pi^2} (\bar s_R \sigma_{\mu \nu} b_L) F^{\mu\nu},
\\
O_{8L}  &=& {g_s \over 4 \pi^2 }~m_b \bar{s}_{L}
\sigma^{\mu\nu} T^a b_{R}~G^a_{\mu\nu}
\\
O_{8R}  &=& {g_s \over 4 \pi^2 }~m_b \bar{s}_{R}
\sigma^{\mu\nu} T^a b_{L}~G^a_{\mu\nu}
,\\
  O_{9L}&=&({\bar s} b)_{L} ({\bar \ell} \ell)_{V},\\
  O_{9R}&=&({\bar s} b)_{R} ({\bar \ell} \ell)_{V},\\
  O_{10L}&=&({\bar s} b)_{L} ({\bar \ell} \ell)_{A},\\
  O_{10R}&=&({\bar s} b)_{R} ({\bar \ell} \ell)_{A},
\eea
where in addition to the SM operators $O_{7L}$, $O_{8L}$, 
$O_{9L}$ and  $O_{10L}$,
we include new operators $O_{7R}$, $O_{8R}$, $O_{9R}$ and $O_{10R}$. 
The new physics effects can originate from
any of the above operators. The operator $O_{7R}$ has already been
introduced in Ref. \cite{kklm}, while the operators $O_{9R}$ and $O_{10R}$ are
newly introduced in this paper. The operator $O_8$ is a chromo-magnetic dipole
operator.

Relegating the details to Ref. \cite{kklm}, we introduce the helicity
amplitudes for the decay
$$B \to K^*(\to K(p_K)+\pi(p_\pi))+ \ell^+(p_+) + \ell^-(p_-)~,$$ 
which can be expressed by
\begin{eqnarray}
H_{+1}^L &=&   
\left (a_L+ c_L\sqrt{\lambda} \right)~, \nonumber \\
H_{-1}^L &=&   
\left(a_L- c_L\sqrt{\lambda} \right)~,  \nonumber \\
H_{0}^L &=&  
 -a_L\frac{ P\cdot L}{ pl}+\frac{ b_L\lambda}{pl}~,  \nonumber \\
H_{+1}^R &=&   
\left (a_R+ c_R\sqrt{\lambda} \right)~, \nonumber \\
H_{-1}^R &=&   
\left(a_R - c_R\sqrt{\lambda} \right)~, \nonumber \\
H_{0}^R &=&  
 -a_R\frac{ P\cdot L}{ pl}+\frac{ b_R\lambda}{pl}~,
\end{eqnarray}
with $P=p_K+p_\pi$, $L=p_+ + p_-$, $p=\sqrt{P^2}$, $l=\sqrt{L^2}$, and
$\lambda=(m_B^2-p^2-l^2)^2/4-p^2l^2$.
And  $a_R,~b_R,~c_R$ and $a_L,~b_L,~c_L$ are given by 
\begin{eqnarray}
a_L &=& -C_{7-}  \left[2 (P\cdot L) g_+ + L^2 (g_+ +g_-) \right]
+\frac{(C_{9-} -C_{10-}) f}{2m_b}L^2 ,
\label{aaa}\\
b_L &=& -2C_{7-} ( g_+  -  L^2 h )-\frac{(C_{9-}-C_{10-}) a_+}{m_b}L^2,\\
c_L &=& -2C_{7+}  g_+ +\frac{(C_{9+}-C_{10+}) g}{m_b}L^2,\label{ccc}
\end{eqnarray}
\begin{eqnarray}
a_R &=& -C_{7-}  \left[2 (P\cdot L) g_+ + L^2 (g_+ +g_-) \right]
+\frac{(C_{9-} + C_{10-}) f}{2m_b}L^2 ,
\label{aaar}\\
b_R &=& -2C_{7-} (  g_+ - L^2 h )-\frac{(C_{9-}+C_{10-}) a_+}{m_b}L^2,\\
c_R &=& -2C_{7+}  g_+ +\frac{(C_{9+}+C_{10+}) g}{m_b}L^2,\label{cccr}
\end{eqnarray}
where the form factors $g$, $g_+$, $g_-$, $f$, $h$ and $a^+$ of 
$B\to K^*$ decay  are defined in Refs. \cite{kklm,ff2,wise}.
We introduced the Wilson coefficients $C_{7-,7+,9-,9+,10-,10+}$ as
\begin{eqnarray}
C_{7-} &=& C_{7R} -C_{7L},~~~C_{7+} = C_{7R} +C_{7L}, \nonumber\\
C_{9-} &=& C_{9R} -C_{9L},~~~C_{9+} = C_{9R} +C_{9L}, \nonumber\\
C_{10-} &=& C_{10R} -C_{10L},~~~C_{10+} = C_{10R} +C_{10L}. \nonumber
\end{eqnarray}

Using the above helicity amplitudes, the angular distribution of  
$B\to K^*(\to K\pi) \ell^+ \ell^-$  is expressed by,
\begin{eqnarray*}
\frac{d^5 \Gamma }{dp^2dl^2 d\cos \theta_ K d\cos \theta_+ d \phi} &=& 
\frac{
\alpha^2 G_F^2 g_{K^*K\pi}^2 \sqrt{\lambda}p^2 m_b^2 |V_{tb} V_{ts}^*|^2}
{64 \times 
8 (2\pi)^8 m_B^3 l^2 [(p^2-m_{K^*}^2)^2 + m_{K^*}^2 \Gamma_{K^*}^2]} 
\end{eqnarray*}
\begin{eqnarray}
~~~~~~~~&\times &\left\{
4 \cos ^2 \theta_K  \sin^2 \theta_+  (|H_{0}^R|^2+|H_{0}^L|^2) \right.
\nonumber \\
&&+\sin^2 \theta_K  (1+ \cos^2 \theta_+ ) ( |H_{+1}^L|^2+  |H_{-1}^L|^2
+|H_{+1}^R|^2+  |H_{-1}^R|^2) \nonumber \\
&&-2\sin ^2\theta_K \sin^2 \theta_+ \left[ \cos 2\phi Re (H_{+1}^R H^{R*}_{-1}+
H_{+1}^L H^{L*}_{-1} ) \right.
\nonumber \\
&&- \sin 2\phi  \left.Im (H_{+1}^R H^{R*}_{-1}+H_{+1}^L H^{L*}_{-1} ) \right]\nonumber \\
&&-\sin 2\theta_K \sin 2\theta_+ \left[\cos \phi Re (H_{+1}^R H^{R*}_{0}
+H_{-1}^RH_0^{R*}+
H_{+1}^L H^{L*}_{0}+H_{-1}^LH_0^{L*})\right.
\nonumber\\
 &&-\sin \phi \left. Im (H_{+1}^R H^{R*}_{0}-H_{-1}^RH_0^{R*}
+H_{+1}^L H^{L*}_{0}-H_{-1}^LH_0^{L*} ) \right] \nonumber \\
&& -2 \sin^2\theta_K \cos\theta_+ 
(|H_{+1}^R|^2-|H_{-1}^R|^2- |H_{+1}^L|^2+|H_{-1}^L|^2) \nn \\
&&+2 \sin\theta_+ \sin 2\theta_K 
 \left [\cos \phi Re (H_{+1}^R H^{R*}_{0}
-H_{-1}^RH_0^{R*}-H_{+1}^L H^{L*}_{0}
+H_{-1}^LH_0^{L*}) \right. \nn \\
&&-  \sin \phi  \left.\left.
Im (H_{+1}^R H^{R*}_{0}+ H_{-1}^R H^{R*}_{0}
- H_{+1}^L H^{L*}_{0}-H_{-1}^L H^{L*}_{0})
\right] \right\} ~.
\eea 
Here we introduced the various angles  as:
$\theta_K$ is the polar   angle of the
$K$ meson momentum in the rest system of the $K^*$
meson with respect to the helicity axis,
{\it i.e.} the outgoing direction of $K^*$.
Similarly $\theta_+$ is the polar angle of
the positron in the $\gamma^*$
rest system with respect to the helicity axis of the $\gamma^*$.
And  $\phi$ is
the azimuthal angle between the planes of the two decays
$K^* \to K\pi$ and $ \gamma^* \to \ell^+ \ell^-$.

If we integrate out the angles $\theta_K$ and $\theta_+$, we get the $\phi$ 
distribution
\begin{eqnarray}
\frac{d \Gamma }{ d \phi} &=& \int
\frac{ 
\alpha^2 G_F^2 g_{K^*K\pi}^2 \sqrt{\lambda}p^2 m_b^2 |V_{tb} V_{ts}^*|^2}
{9\times 16 (2\pi)^8 m_B^3 l^2 [(p^2-m_{K^*}^2)^2 + m_{K^*}^2 \Gamma_{K^*}^2]} 
\left\{
  |H_{0}^R|^2 + |H_{+1}^R|^2+  |H_{-1}^R|^2 \right.\nonumber \\
&&|H_{0}^L|^2 + |H_{+1}^L|^2+  |H_{-1}^L|^2 
-\cos 2\phi Re (H^R_{+1} H^{R*}_{-1}+H_{+1}^L H^{L*}_{-1} )\nonumber \\
&&\left.+ \sin 2\phi Im (H_{+1}^R H^{R*}_{-1}+H^L_{+1} H^{L*}_{-1} )
 \right\} d p^2 d l^2~.
\label{eqphi}
\end{eqnarray} 
Similarly, we can get the $\theta_K$ and $\theta_+$ angular distributions 
as following: 
\begin{eqnarray}
\frac{d \Gamma }{ d \cos\theta_K} &=& \int
\frac{(2 \pi)
\alpha^2 G_F^2 g_{K^*K\pi}^2 \sqrt{\lambda}p^2 m_b^2 |V_{tb} V_{ts}^*|^2}
{3\times 64 (2\pi)^8 m_B^3 l^2 [(p^2-m_{K^*}^2)^2 + m_{K^*}^2 \Gamma_{K^*}^2]} 
\left\{ 2 \cos^2 \theta_K (
  |H_{0}^R|^2 \right.\nonumber \\
&&+\left. |H_{0}^L|^2)+ \sin^2\theta_K \left(|H_{+1}^R|^2+  |H_{-1}^R|^2+
|H_{+1}^L|^2+  |H_{-1}^L|^2 \right)
 \right\} d p^2 d l^2~,
\label{eqtheta}
\end{eqnarray}
and
\begin{eqnarray}
\frac{d \Gamma }{ d \cos\theta_+} &=& \int
\frac{(2 \pi)
\alpha^2 G_F^2 g_{K^*K\pi}^2 \sqrt{\lambda}p^2 m_b^2 |V_{tb} V_{ts}^*|^2}
{6\times 64 (2\pi)^8 m_B^3 l^2 [(p^2-m_{K^*}^2)^2 + m_{K^*}^2 \Gamma_{K^*}^2]} 
\left\{ 2 \sin^2 \theta_+ ( |H_{0}^R|^2  + |H_{0}^L|^2) \right. \nonumber \\ 
&&+\left. (1 + \cos\theta_+)^2 \left(|H_{+1}^L|^2+  |H_{-1}^R|^2 \right) +
(1 - \cos\theta_+)^2 \left(|H_{+1}^R|^2+  |H_{-1}^L|^2 \right)
 \right\} d p^2 d l^2~.
\label{eqthetap}
\end{eqnarray}

Taking the narrow resonance limit of $K^*$ meson, {\it i.e.}
using the  equations
\begin{eqnarray}
&&\Gamma_{K^*}=\frac{g_{K^*K\pi}^2 m_{K^*}}{48 \pi}~, \nn \\
&&\lim_{\Gamma_{K^*} \to 0} 
\frac{\Gamma_{K^*} m_{K^*}}{(p^2-m_{K^*}^2)^2 + m_{K^*}^2 \Gamma_{K^*}^2}=
\pi \delta(p^2-m_{K^*}^2)~,
\end{eqnarray}
we can perform the integration over $p^2$ and obtain the
double differential branching fraction with respect to 
dilepton mass squared $l^2$ and angle variables, 
\begin{eqnarray}
{dBr \over dl^2 d \phi}&=& \tau_{B}{\alpha^2 {G_F}^2 \over 384 \pi^5}
\sqrt{\lambda} {{m_b}^2 \over {m_B}^3 l^2} |V_{ts}V_{tb}|^2 {1 \over 2 \pi}
\left\{
  |H_{0}^R|^2 + |H_{+1}^R|^2+  |H_{-1}^R|^2 \right.\nonumber \\
&&+|H_{0}^L|^2 + |H_{+1}^L|^2+  |H_{-1}^L|^2 
-\cos 2\phi Re (H^R_{+1} H^{R*}_{-1}+H_{+1}^L H^{L*}_{-1} )\nonumber \\
&&\left.+ \sin 2\phi Im (H_{+1}^R H^{R*}_{-1}+H^L_{+1} H^{L*}_{-1} )
 \right\}~, 
\label{lphi}
\end{eqnarray}

\begin{eqnarray}
{dBr \over dl^2 d \cos \theta_K}&=& \tau_{B}{\alpha^2 {G_F}^2 \over 384 \pi^5}
\sqrt{\lambda} {{m_b}^2 \over {m_B}^3 l^2} |V_{ts}V_{tb}|^2 {3 \over 4}
\left\{
 2 \cos^2 \theta_K (|H_{0}^R|^2 + |H_{0}^L|^2)\right.\nonumber \\
&&\left. + \sin^2 \theta_K (|H_{+1}^R|^2 + |H_{-1}^R|^2 + |H_{+1}^L|^2 
+ |H_{-1}^L|^2)  \right\}~, 
\label{lthetak}
\end{eqnarray}

\begin{eqnarray}
{dBr \over dl^2 d \cos \theta_+}&=& \tau_{B}{\alpha^2 {G_F}^2 \over 384 \pi^5}
\sqrt{\lambda} {{m_b}^2 \over {m_B}^3 l^2} |V_{ts}V_{tb}|^2 {3 \over 8}
\left\{
 2 \sin^2 \theta_+ (|H_{0}^R|^2 + |H_{0}^L|^2)\right.\nonumber \\
&&\left. + (1+\cos \theta_+)^2 (|H_{+1}^L|^2 + |H_{-1}^R|^2)
+ (1-\cos \theta_+)^2 (|H_{-1}^L|^2 + |H_{+1}^R|^2)  \right\}~, 
\label{lthetap}
\end{eqnarray}
and
\begin{eqnarray}
 {dBr \over d{l^2}}&=&\tau_{B}{\alpha^2 {G_F}^2 \over 384 \pi^5}
\sqrt{\lambda} {{m_b}^2 \over {m_B}^3 l^2} |V_{ts}V_{tb}|^2
\left\{
  |H_{0}^R|^2 + |H_{+1}^R|^2+  |H_{-1}^R|^2 \right.\nonumber \\
 &&\left. +|H_{0}^L|^2 + |H_{+1}^L|^2+  |H_{-1}^L|^2 \right\}~, 
\label{38}
\end{eqnarray}
where $\tau_{B}$ is the life time of $B$ meson, and 
we replace all $p$ by $m_{K^*}$ due to the
$\delta$ function. 

Finally, to eliminate the constant factors in Eq.  (\ref{lphi}), we define
the  normalized  distribution,
\begin{eqnarray}
r(\phi, \hat{s})
&\equiv& \left[{dBr \over d{l^2} d \phi}\right]/
\left[{dBr \over d{l^2} }\right] \nn \\
&=&{1 \over 2 \pi} \left\{
1-\frac{\cos 2\phi Re (H^R_{+1} H^{R*}_{-1}+H_{+1}^L H^{L*}_{-1} )
- \sin 2\phi Im (H_{+1}^R H^{R*}_{-1}+H^L_{+1} H^{L*}_{-1} )}
{|H_{0}^R|^2 + |H_{+1}^R|^2+  |H_{-1}^R|^2 
 +|H_{0}^L|^2 + |H_{+1}^L|^2+  |H_{-1}^L|^2} \right\}, 
\label{R}
\end{eqnarray}
where $\hat{s}=l^2 /m_B^2$. The distribution
$r(\phi,\hat{s})$ is the probability for finding $K$ meson per unit 
radian region in the direction of azimuthal angle $\phi$.
Therefore, the normalized distribution $r(\phi,\hat{s})$ oscillates 
around its average value given by ${1 \over 2 \pi}\simeq 0.16$. 

\section{Gluino Mediated Flavor Changing Neutral Current and 
its Numerical Analyses}

In this section we consider the flavor changing neutral current (FCNC) in
generalized supersymmetric models (gSUSYs).  In gSUSYs, the soft mass
terms for sfermions can  lead to potentially large FCNC \cite{gabbiani}.
In the mass-insertion-approximation (MIA) \cite{hall}, one chooses a basis for
fermion and sfermion states, in which all the couplings of these particles
to neutral gauginos are flavor diagonal. Flavor changes in the squark sector
are provided by the non-diagonality of the sfermion propagators,
which can be expressed in terms of 
the dimensionless parameters $(\delta_{ij}^q)_{MN}$,
\begin{eqnarray}
(\delta_{ij}^q)_{MN} = \frac{(m_{ij}^{\tilde q})^2_{MN}}{\tilde m}
~~~~(M, N = L, R)~,
\end{eqnarray}
where $(m_{ij}^{\tilde q})^2_{MN}$ are the off-diagonal elements 
of the $\tilde q$ mass squared matrix that mixes flavor $i,j$ for
both left- and right-handed scalars, and $\tilde m$ is the average 
squark mass. 
The expressions for the Wilson coefficients at $M_W$ scale due to the FCNC
gluino exchange diagrams \cite{gabbiani} are
\begin{eqnarray}
C_{7L}^{\rm SUSY} (M_W)
& = & {8 \pi \alpha_s  \over 9 \sqrt{2} G_F  \tilde{m}^2
\lambda_t} ~\left[ \left( \delta_{23}^d \right)_{LL} M_3 (x) +
\left( \delta_{23}^d \right)_{LR} {m_{\tilde{g}} \over m_b} M_1 (x) \right]~ ,
\nonumber
\\
C_{8L}^{\rm SUSY} ( M_W)
& = & { \pi \alpha_s  \over  \sqrt{2} G_F  \tilde{m}^2
\lambda_t}~\left[ \left( \delta_{23}^d \right)_{LL} \left( {1\over 3} M_3 (x)
+ 3 M_4 (x) \right) +  \left( \delta_{23}^d \right)_{LR} {m_{\tilde{g}} \over
m_b} \left( {1\over 3} M_1 (x) + 3 M_2 (x) \right) \right],
\nonumber
\\
C_{9L}^{\rm SUSY} ( M_W ) & = &
{16 \pi \alpha_s  \over 9 \sqrt{2}  G_F  \tilde{m}^2
\lambda_t} ~\left( \delta_{23}^d \right)_{LL} P_1 (x) ~,
\label{26}
\end{eqnarray}
and
\begin{eqnarray}
C_{7R}^{\rm SUSY} (M_W)
& = & {8 \pi \alpha_s  \over 9 \sqrt{2} G_F  \tilde{m}^2
\lambda_t} ~\left[ \left( \delta_{23}^d \right)_{RR} M_3 (x) +
\left( \delta_{23}^d \right)_{RL} {m_{\tilde{g}} \over m_b} M_1 (x) \right]~ ,
\nonumber
\\
C_{8R}^{\rm SUSY} ( M_W)
& = & { \pi \alpha_s  \over  \sqrt{2} G_F  \tilde{m}^2
\lambda_t}~\left[ \left( \delta_{23}^d \right)_{RR} \left( {1\over 3} M_3 (x)
+ 3 M_4 (x) \right) +  \left( \delta_{23}^d \right)_{RL} {m_{\tilde{g}} \over
m_b} \left( {1\over 3} M_1 (x) + 3 M_2 (x) \right) \right],
\nonumber
\\
C_{9R}^{\rm SUSY} ( M_W ) & = &
{16 \pi \alpha_s  \over 9 \sqrt{2}  G_F  \tilde{m}^2
\lambda_t} ~\left( \delta_{23}^d \right)_{RR} P_1 (x)~, 
\label{27}
\end{eqnarray}
with $\lambda_t \equiv V_{ts}^* V_{tb}$.
The functions $M_{1,3}(x)$ and $P_1 (x)$ are defined as 
\begin{eqnarray}
M_1 (x) & = & {1 + 4 x - 5x^2 + ( 4 x + 2 x^2 ) \ln x \over 2 (1-x)^4}~,
\nonumber   \\
M_2 (x) & = & -x^2 { 5 - 4 x - x^2 + ( 2 + 4 x) \ln x \over
2 ( 1 - x)^4 }~,
\nonumber  \\
M_3 (x) & = & {-1 + 9 x + 9 x^2 - 17 x^3 + ( 18 x^2 + 6 x^3 ) \ln x
\over 12 (x-1)^5}~,
\nonumber   \\
M_4 (x) & = & {-1 - 9 x + 9 x^2 + x^3 - 6 (x + x^2 ) \ln x \over
6 ( x - 1)^5}~,
\nonumber   \\
P_1 (x) & = & { 1 - 6 x + 18 x^2 - 10 x^3 - 3 x^4 + 12 x^3 \ln x
\over 18 (x-1)^5 }~,
\end{eqnarray}
where $x \equiv m^2_{\tilde g}/{\tilde m}^2$ and $m_{\tilde g}$ is
gluino mass.

In addition to the above gSUSYs contributions, the usual
SM contributions $C_{7L}^{\rm SM}$, $C_{8L}^{\rm SM}$,
 $C_{9L} ^{\rm SM}$, and $C_{10L}^{\rm SM}$ are already known for years,
which we will not show here. Please look at Refs. \cite{kklm,ab} for details
of the SM contributions.
Including the QCD corrections, we get the Wilson coefficients at
$m_b$ scale as 
\begin{eqnarray}
C_{7L}(m_b) &=& -0.31 + 0.67 C_{7L}^{\rm SUSY}(M_W) 
+ 0.09 C_{8L}^{\rm SUSY}(M_W)~, \label{eq31}
\\
C_{7R}(m_b) &=& 0.67 C_{7R}^{\rm SUSY}(M_W) + 0.09 C_{8R}^{\rm SUSY}(M_W)~,
\\
C_{9L} (m_b) &=& C_{9L}^{\rm SM}(m_b) +  C_{9L}^{\rm SUSY}(M_W)~,
\\
C_{9R} (m_b) &=& C_{9R}^{\rm SUSY}(M_W)~.
\end{eqnarray}
Here, $-0.31$ in Eq. (\ref{eq31}) is the SM value of $C_{7L}(m_b)$. New physics
contributions $C_{iL}^{\rm SUSY}(M_W)$, $C_{iR}^{\rm SUSY}(M_W)$
 come from Eqs. (\ref{26},\ref{27}).
For $C_{10}$, there is no new physics contribution,
\begin{eqnarray}
C_{10L} (m_b) &=& C_{10L}^{\rm SM} (m_b)~,
\\
C_{10R} (m_b) &=& 0~.
\end{eqnarray}
Then we can get the complete formula for angular distributions of 
$B\to K^* \ell^+\ell^-$, Eq. (\ref{R}).
 
The operators $O_{7L}$ and $O_{7R}$ contribute to the rare radiative decay
$b\to s \gamma$. 
Their Wilson coefficients have been constrained by the experimental
measurements of the decay.
The decay width for inclusive $b \to  s \gamma$ decay is given 
in terms of the operators $O_{7L}$ and $O_{7R}$.
It is convenient to normalize this radiative partial width to
the semileptonic decay
$b \to  c e \bar{\nu}$
in terms of the ratio $R$, 
\begin{equation}
R \equiv {\Gamma (b \to  s \gamma) 
\over \Gamma(b \to  c e \bar{\nu})} =
{6 \over \pi} {|V_{ts}^* V_{tb}|^2 \over |V_{cb}|^2}
{\alpha_{em} \over f({m_c / m_b})}
{ |C_{7L} (m_b)|^2 + |C_{7R} (m_b)|^2 \over 
1 - {2 \over 3 \pi} \alpha_s (m_b) g({m_c / m_b}) } ,\label{rr}
\end{equation}
where the functions $f(x)$ and $g(x)$ are phase space and QCD correction 
factors  \cite{12}, respectively.  The $b \to  s \gamma$ branching fraction is
obtained by 
\begin{equation}
{\cal BR}(b \to  s \gamma) 
\simeq {\cal BR}(B \to X_c l \nu)_{\rm exp.} \times R
\simeq (0.105) \times R .
\end{equation}
For ${\cal BR}(b \to  s \gamma)$, 
we use the present experimental value \cite{cleo} of the branching 
fraction for the inclusive $B \to  X_s \gamma$ decay, 
\begin{equation}
{\cal BR}(B \to  X_s \gamma) = 
(3.15 \pm  0.35 \pm 0.32 \pm 0.26) \times 10^{-4} .
\end{equation}
Constrained by this experiment, we derive from Eq. (\ref{rr})
\begin{equation}
|C_{7L} (m_b)|^2 + |C_{7R} (m_b)|^2=0.081\pm 0.014.\label{c7}
\label{constraint}
\end{equation}
In the numerical calculations, we use
the form factors calculated in Ref. \cite{ff2}. 
They are listed in Table~ \ref{t1} for zero momentum transfer. 
The evolution formula for these form factors is 
\begin{equation}
f_i (l^2) = \frac{f_i (0)}{ 1- \sigma_1 l^2 +\sigma_2 l^4}~,
\end{equation} 
where $l^2 = (p_{\ell^+} + p_{\ell^-})^2$.
The corresponding values $\sigma_1$ and $\sigma_2$ for each form factors 
are also listed in Table 1.

The decay $B\to K^* + J/\psi ( \to \ell^+ \ell^-)$ is a possible background for
our $B\to K^* \ell^+\ell^-$ decay at the $J/\psi$ resonance region, so as
the $\psi '$, $etc$. 
Therefore, only the low invariant mass region of the lepton pair is good
for clean measurements. 
The helicity amplitudes are dominated by 
the two coefficients $C_{7L}$ and $C_{7R}$ in the region of low invariant 
mass,  as given by
\begin{eqnarray}
H^{L,R}_{+1}& \simeq&-4 g_+ C_{7R} \sqrt{\lambda}~, \nn \\
H^{L,R}_{-1}& \simeq& 4 g_+ C_{7L} \sqrt{\lambda}~,\nn \\
H^{L,R}_{0}& \simeq &0~.
\end{eqnarray}
In the small invariant mass 
limit $\hat s \ll 1$, $r(\phi,\hat{s})$, defined in Eq. (\ref{R}),
is approximately written as,
\begin{eqnarray}
r(\phi,~ \hat{s} \ll 1) &\simeq&{1 \over 2 \pi} \left\{
1+\cos 2\phi \frac{Re (C_{7R} C_{7L}^*)}{|C_{7R}|^2+|C_{7L}|^2}
- \sin 2\phi \frac{Im (C_{7R} C_{7L}^*)}{|C_{7R}|^2+|C_{7L}|^2}
\right\} .
\label{approx}
\end{eqnarray}
In the SM case, $C_{7R} \simeq 0$ and therefore the above approximate 
formula is reduced to
\begin{eqnarray}
r(\phi,~ \hat{s} \to 0)_{\rm SM} &\simeq&{1 \over 2 \pi} .
\end{eqnarray}
In Fig. 1 we can see that it is almost a constant
 distribution of $\phi$ in the small $\hat s$ region.
As $\hat s$ increases, the contributions from the operator 
$O_9$ and $O_{10}$ makes  $(\sim -\cos 2\phi)$ behavior.
However, the new physics contributions can give quite different 
distributions depending on the model, and we can probe new physics efficiently.
Here we discuss the gSUSYs contribution to the distribution $r(\phi,\hat{s})$.
For simplicity, we assume that $|\delta_{LR}|=|\delta_{RL}|$ and 
$|\delta_{LL}|=|\delta_{RR}|$, and consider two cases: 
LR mixing dominating case and LL mixing dominating case.

First we consider 
the LR mixing dominating situation, $i.e.$ 
$|\delta_{LL}| \sim |\delta_{RR}|  \ll |\delta_{LR}| \sim |\delta_{RL}|$.
Fig.~\ref{f1} shows the distribution $r(\phi, \hat{s})$ 
for $\delta_{LR}=\delta_{RL}=|\lambda_t|$ case
with $x=0.3$, $\tilde m = 960$ GeV. 
This corresponds to $C_{7L}=0.017$ and $C_{7R}=0.333$.
Since  $C_{7L} \ll C_{7R}$ and both are real, 
the approximate formula (\ref{approx}) becomes
\begin{equation}
r(\phi,~\hat{s} \ll 1) \simeq {1 \over 2 \pi}\left\{1 +
 \cos 2\phi \left |\frac{ C_{7L}}{C_{7R}}\right| \right\}.
\end{equation}
This $(\sim \cos 2\phi)$ behavior is shown in Fig.~\ref{f1}.
On the other hand,
Fig.~\ref{f2} shows the distribution $r(\phi, \hat{s})$ 
for $\delta_{LR}=-\delta_{RL}=|\lambda_t|$ case
with $x=0.3$, $\tilde m = 960$ GeV. 
This corresponds to $C_{7L}=0.017$ and $C_{7R}=-0.333$.
In this case the approximate formula (\ref{approx}) becomes
\begin{equation}
r(\phi,~\hat{s} \ll 1) \simeq {1 \over 2 \pi}\left\{1 -
 \cos 2\phi \left |\frac{ C_{7L}}{C_{7R}}\right| \right\}.
\end{equation}
This $(\sim -\cos 2\phi)$ behavior is shown in Fig.~\ref{f2} explicitly
for the small $\hat s$ region.
In Figs.~\ref{f1} and ~\ref{f2}, 
we used the same values of $x$ and $\tilde m$:
The branching fractionss of both $b\to s \gamma$ and $b\to s \ell^+\ell^-$ are
unchanged for the two situations, and we cannot separate these two situations
by using only branching fraction measurements. 
However, we can see that the angular distributions, shown in Figs. 2
and 3, can  easily distinguish the relative sign of $\delta_{LR}$ and
$\delta_{RL}$.

For the LL mixing dominating case, $i.e.$ 
$|\delta_{LR}| \sim |\delta_{RL}| \ll |\delta_{LL} |\sim |\delta_{RR}|$, 
we also show two cases.
First we choose 
$\delta_{LL}=\delta_{RR}=e^{i\frac{\pi}{4}}$,
with $x=0.8$, $\tilde m = 250$ GeV. 
This corresponds to $C_{7L}=-0.16 + i~0.15$ and 
$C_{7R}=+0.15 + i~0.15$, $i.e.$
${C_{7R}}/{C_{7L}} \sim e^{i\frac{3\pi}{2}}$ case.
Using this set of parameters, the formula (\ref{approx})
becomes 
\begin{eqnarray}
r(\phi,~\hat{s} \ll 1) &\simeq&{1 \over 2 \pi} \left\{
1+\frac{1}{2} \sin 2\phi 
\right\} .
\end{eqnarray}
Fig. \ref{f3} shows this  $(\sim \sin 2\phi)$ behavior clearly.
On the other hand,
Fig. 5 shows the distribution $r(\phi, \hat{s})$ 
for $\delta_{LL}=-\delta_{RR}=e^{i\frac{\pi}{4}}$ case
with $x=0.8$, $\tilde m = 250$ GeV. 
This corresponds to $C_{7L}=-0.16 + i~0.15$ and 
$C_{7R}=-0.15 - i~0.15$, i.e,
${C_{7R}}/{C_{7L}} \sim e^{i\frac{\pi}{2}}$ case. 
The approximate formula becomes
\begin{eqnarray}
r(\phi,~\hat{s} \ll 1) &\simeq&{1 \over 2 \pi} \left\{
1 - \frac{1}{2} \sin 2\phi 
\right\} .
\end{eqnarray}
The $(\sim -\sin 2\phi)$ behavior is shown in Fig. 5.
From Figs. 4 and 5, we can see that the angular distribution can distinguish
the relative phase between $\delta_{LL}$ and $\delta_{RR}$ easily, 
even if we use the same values of parameters, $x$ and $\tilde m$.

The polar angle distribution functions in Eqs. (\ref{lthetak},\ref{lthetap})
depend  only on the modular square terms of the helicity amplitudes, which
give the  decay width of the semileptonic decay. 
If the branching fraction is
fixed by experiments, these two angle distributions cannot distinguish new
physics contribution from the SM. 
On the other hand, they can serve as a double check of whether the branching
fraction is different from the SM predictions.


In conclusion, we have calculated the angular distribution of the rare decay,
$B\to K^*+ ( \to K \pi) + \ell^+ + \ell^-$, in general supersymmetric
extensions of the standard model.  The azimuthal angle ($\phi$) distribution
in gSUSYs
can be quite different from that of the SM,  
while the measured branching fraction for $B \to X_s \gamma$ can be
accommodated within the standard model prediction.
In the standard model it is found to be almost a constant 
under the variation of the angle $\phi$ in small invariant mass region,
while in gSUSYs the distribution can show $(\sim \pm \cos 2\phi)$ or
$(\pm \sin 2\phi)$  behavior depending on the gSUSYs parameters. 
We showed that the angular distribution of the
decay can tell us the new physics effects clearly from the behavior of the
distribution, even if new physics does not change the decay rate
substantially:   We would be able to tell the
relative phase between the mixing parameters $\delta_{LR}$ and $\delta_{RL}$ 
(or  $\delta_{LL}$ and $\delta_{RR}$), even though the decay rate of gSUSYs
were exactly the same as that of the SM.
\\

\section*{Acknowledgement}

We thank G. Cvetic and T. Morozumi for careful reading of the manuscript 
and their valuable comments.
The work of C.S.K. was supported in part by Seo-Am (SBS) Foundation, 
in part by  BK21 Project, SRC Program and Grant No. 2000-1-11100-003-1
of the KOSEF, and in part by the KRF Grants, Project No. 2000-015-DP0077.
The work of Y.G.K. is supported by JSPS.


\begin{table}[hb]
\caption{Form factors in zero momentum transfer and parameters of
evolution formula [11].}
\begin{center}
\begin{tabular}{c|ccccccc}
\hline
            & $g$  &  $f$ &  $a_+$  & $a_-$  &  $g_+$  & $g_-$  &  $h$ \\
$f_i(0)$    &0.063 & 2.01 & -0.0454 & 0.053  & -0.3540 & 0.313  & -0.0028 \\
\hline
$\sigma_1$ & 0.0523  & 0.0212  & 0.039 & 0.044 & 0.0523 & 0.053 & 0.0657\\
$\sigma_2$ & 0.00066 & 0.00009 &0.00004&0.00023& 0.0007 &0.00067& 0.0010\\
\hline
\end{tabular}\end{center}\label{t1}
\end{table}

\begin{figure}\begin{center}
\epsfig{file=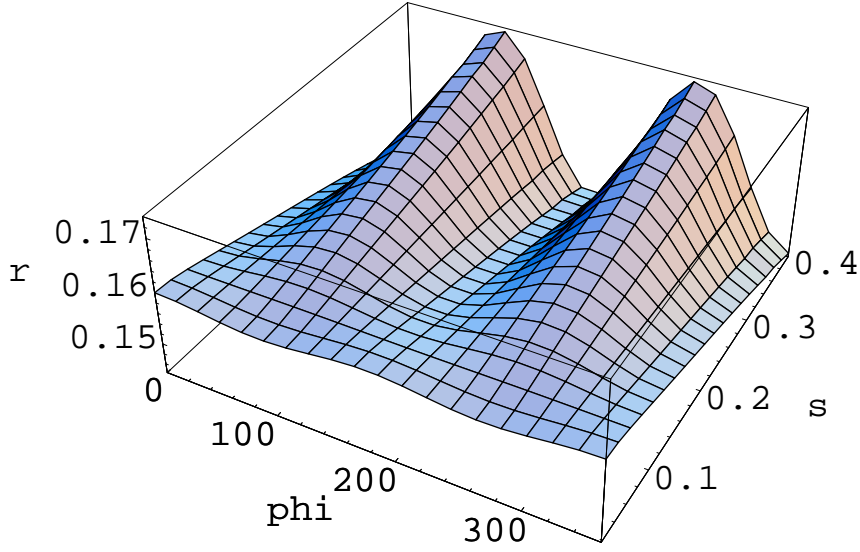, 
width=11.5cm,angle=0}
\caption{The distribution $r(\phi, \hat{s})$ 
for the SM case, where $C_{7R}\simeq 0$.} \label{f0}
\end{center}
\end{figure}

\begin{figure}\begin{center}
\epsfig{file=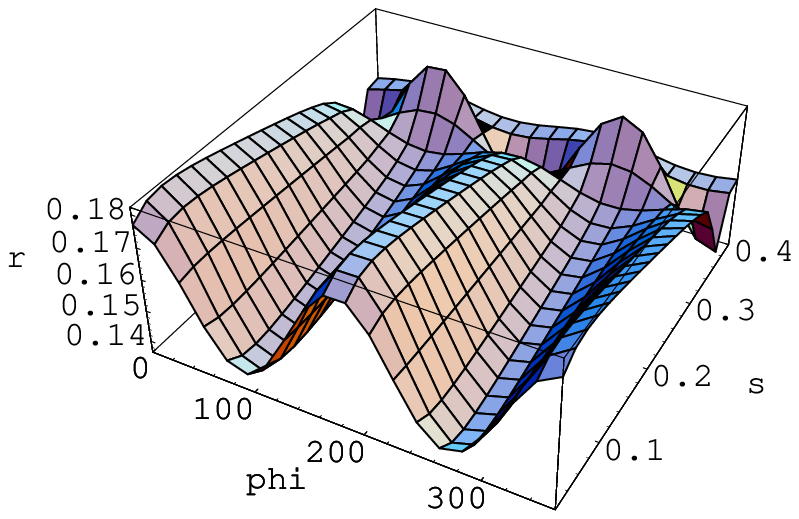, 
width=11.5cm,angle=0}
\caption{The distribution $r(\phi, \hat{s})$ 
for $\delta_{LR}=\delta_{RL}=|\lambda_t|$ case,
with $x=0.3$, $\tilde m = 960$ GeV.
This corresponds to $C_{7L}=0.017$, $C_{7R}=0.333$.} \label{f1}
\end{center}
\end{figure}

\begin{figure}\begin{center}
\epsfig{file=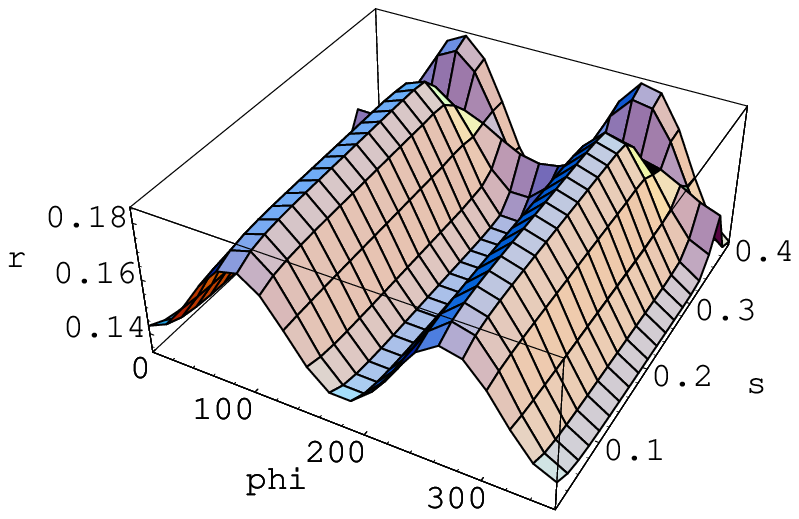, 
width=11.5cm,angle=0}
\caption{The distribution $r(\phi, \hat{s})$ 
for $\delta_{LR}=-\delta_{RL}=|\lambda_t|$ case,
with $x=0.3$, $\tilde m = 960$ GeV.
This corresponds to $C_{7L}=0.017$, $C_{7R}=-0.333$.} \label{f2}
\end{center}
\end{figure}

\begin{figure}\begin{center}
\epsfig{file=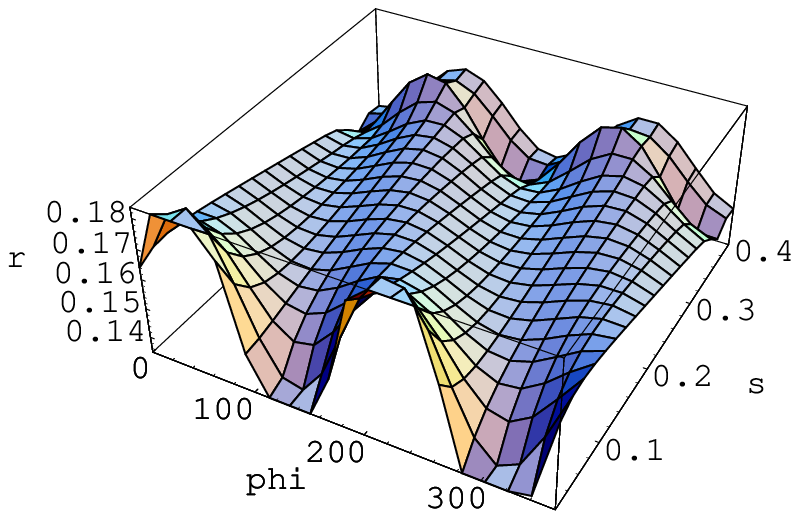, 
width=11.5cm,angle=0}
\caption{The distribution $r(\phi, \hat{s})$ 
for $\delta_{LL}=\delta_{RR}=e^{i\frac{\pi}{4}}$ case,
with $x=0.8$, $\tilde m = 250$ GeV.
In this case, $C_{7R}/C_{7L}\simeq e^{i\frac{3}{2}\pi}$.} \label{f3}
\end{center}
\end{figure}

\begin{figure}\begin{center}
\epsfig{file=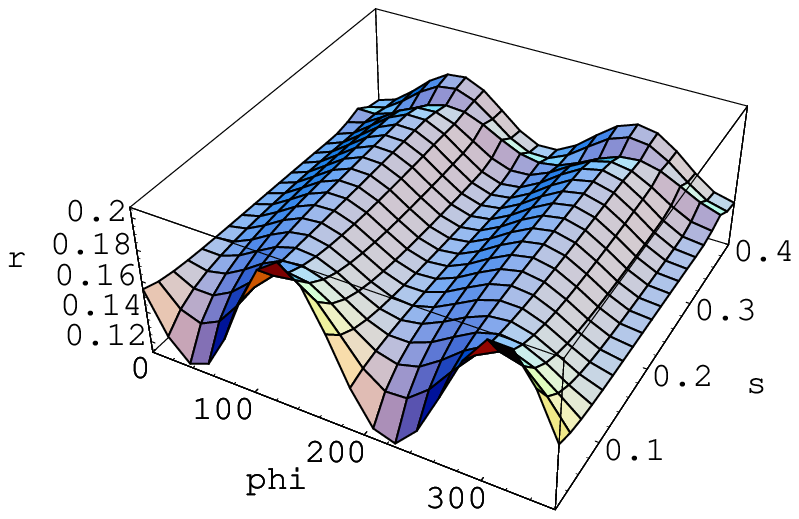, 
width=11.5cm,angle=0}
\caption{The distribution $r(\phi, \hat{s})$ 
for $\delta_{LL}=-\delta_{RR}=e^{i\frac{\pi}{4}}$ case,
with $x=0.8$, $\tilde m = 250$ GeV.
In this case, $C_{7R}/C_{7L}\simeq e^{i\frac{\pi}{2}}$.} \label{f4}
\end{center}
\end{figure}

\end{document}